
\magnification=\magstep1
\baselineskip=15pt
\centerline{\bf  Constraints and Hamiltonian in Light-front Quantized Field
Theory}
\vskip 1cm
\centerline {Prem P. Srivastava\footnote* {Permanent address:
{\it Centro Brasileiro de Pesquisas F\'isicas, Rua Xavier Sigaud 150,
22290 Rio de Janeiro, R.J. Brasil.} Bitnet: prem@cbpfsu1.cat.cbpf.br  }}

\vskip 0.4cm
\centerline {{\it INFN} and {\it Dipartimento di Fisica}
Galileo Galilei, {\it Universit\`a di Padova, Padova,}}

\centerline {and}

\centerline {\it Department of Physics,
The Ohio State University, Columbus, Ohio. }

\vskip 0.5cm

\centerline {\bf Abstract}
\vskip 0.5cm

\baselineskip=15pt

Self-consistent Hamiltonian formulation of scalar theory on the null
plane is constructed following Dirac method.
The theory contains also {\it constraint equations}. They would give,
if solved, to a nonlinear and nonlocal Hamiltonian. The
constraints lead us in the continuum to a different description of
spontaneous symmetry breaking since,
the symmetry generators now annihilate the vacuum. In two examples
where the procedure lacks self-consistency,
the corresponding theories are known ill-defined from equal-time
quantization. This lends support to the method adopted where
both the background field and the fluctuation above it are
treated as dynamical variables on the null plane. We let the
self-consistency of the Dirac procedure determine their properties
in the quantized theory. The results following
from the continuum and the
discretized formulations in the infinite volume limit do agree.

\vskip 2.5cm
\noindent DFPF/92/TH/58 \par
\noindent December 1992\par

\vfill
\eject

\baselineskip=15pt

\noindent {\bf 1. Introduction:}

The possibility of building hamiltonian
formulation of relativistic
dynamics on light-front surface, $(t+z)={\it const.}$, was
indicated by Dirac [1] and rediscovered [2]
in the context of the infinite-momentum frame.
The longitudinal momentum $k^+$ (say, in the massive case)
being necessarily positive and conserved, the presence of $k^+=0$
quanta in the light-front vacuum seems excluded. This implies a simpler
ground state compared to that encountered in the case of
equal-time quantization.
The much discussed discretized light-cone quantized
(DLCQ) field theory [3] does
reveal significant simplifications. Among the  recent developments is the
Light-front Tamm-Dancoff Field Theory which has been proposed for
the non-perturbative bound state problem [4,3].

However, the description on the null-plane of
the spontaneous symmetry breaking (ssb) or Higgs mechanism,
for example, are not clearly understood even at the tree level [3,4].
We will address to it by quantizing the self-interacting scalar field.
For this we need first to decide
which Hamiltonian to use. The ingredients commonly employed,
e.g., the (perturbative) vacuum,
the canonical commutation relations,
and the local Hamiltonian with polynomial self-interaction terms
are perhaps too simple
to give rise to all the non-perturbative effects,
if we consider that we are faced with a
constrained dynamical system.  Among other motivations to clarify
the problem  are the string theories (e.g., heterotic strings) where
the scalar fields also appear and the null plane quantization frequently
adopted for a better physical understanding.
The Dirac method is a systematic procedure to construct the
hamiltonian formulation for constrained lagrangians.
We show  that a self-consistent
formulation in our context
does exist in the continuum (excepting some special cases)
and that it contains not only
the Hamiltonian but also {\it constraint
equations}. Solving the
constraints would lead to a nonlinear and nonlocal Hamiltonian.
A description of the ssb parallel to that in the equal-time case,
though with some important differences is shown to follow.
When other fields are also present
the constraint eqs. would relate the various vacuum
condensates and may also be useful in  suggesting
new counter terms in the quantized theory.
The Dirac method
was attempted earlier using (the finite volume) discretized formulation [6,7]
in incomplete fashion.
The {\it standard} method requires that {\it all}
the constraints (not associated with the gauge-fixing)
be derived from inside the given Lagrangian.
In  [6] the constraint $p\approx 0$ below was missed or the
method not completed by finding the Dirac brackets in order to
implement the constraints while
in [7] constraints were added from outside. The check of the self-consistency
of the procedure [5] was overlooked and
the properties of the symmetry generators required to
describe the ssb were not considered.
It is also pointed out that the discussion may be made in
the continuum or in the finite volume with a convenient boundary condition;
they agree when the infinite volume limit of the latter is taken
(e.g. ref. [8]).
The physically significant predictions are restored
only by removing the  spurious
finite size effects corresponding to, say,
the zero modes in the finite volume sgn
and delta functions. Two examples (Secs. 2,3) are given
where Dirac procedure becomes inconsistent; the
corresponding theories, however, are known from the equal-time
quantization studies to be ill-defined.
This lends support to
the method followed here requiring a careful self-consistency check.
In Sec. 2 we explain the details for scalar field
in two dimensions. The extensions
to an iso-multiplet in $3+1$
dimensions is discussed in Sec. 3 where also the symmetry
generators in the quantized theory
needed to give a  description of the ssb (and Higgs mechanism)
are constructed.

\bigskip

\noindent {\bf 2. Hamiltonian Formulation ( Dirac procedure):}

Consider first the case of a real massive
scalar field $\,\phi\,$ in two space-time
dimensions with the Lagrangian density $\, [{\dot\phi}{\phi^\prime}
-V(\phi)],\,$ where an overdot and
a prime indicate the partial derivations with respect to the light-front
coordinates $\,\tau=(x^0+x^1)/{\sqrt2}\,$
and $\,x =(x^0-x^1)/{\sqrt2}\,$ respectively. The eq. of motion
$\,2\dot{\phi'}= -V'(\phi)\;$, where a prime on V indicates
the variational derivative with respect to $\phi$,
shows that the solutions, $\phi=const.$, are possible. If we integrate the
eq. of motion over $\,-L/2\le x\le L/2\,$ we are led to the
constraint equation, $\int dx\,V'(\phi)=
-2\,\partial_{\tau}[{\cal C}(\tau,L)]\,$, where $\,{\cal C}(\tau,L)=
[\,\phi(\tau,x=L/2)-\phi(\tau,x=-L/2)\,]\,$. The constraint thus seems
to depend on the value of the surface variable
${\cal C}(\tau,L)$ which is absent
for the periodic boundary condition but may be
non-vanishing for other cases.  We are thus required to formulate
the (physical) problem at hand more carefully.
Based on the consideration that the possible
constant solutions for $\phi$ at the classical level
are relevant for describing the ground state and the
ssb we should make the separation, as is usually done in the context of
gauge theories,
$\;\phi(x,\tau)=\omega(\tau)+\varphi(x,\tau)\;$. The variable
$\omega \,$ corresponds to the background field (bosonic condensate)
while $\,\varphi\,$ describes the (quantum) fluctuations above the former.
This separation should be done independent of whether we work in the
finite volume (discretized formulation) {\it or} directly in the
continuum formulation. The Lagrangian now reads as

$${\cal C}{\dot\omega} +\int_{-L/2}^{L/2} dx\,{\dot\varphi}\,{\varphi'}
-\int_{-L/2}^{L/2}dx \,V(\phi)\,,\eqno(1)$$

\noindent where the surface variable ${\cal C}(\tau,L)$ is to be treated as a
dynamical one like $\omega $ and $\varphi$ variables.
The set of {\it three} lagrange eqs. then leads to the following
{\it constraint equation}, which is independent of the boundary conditions
adopted,

$$\eqalign {{1\over L}\beta(\tau) &\equiv
{1\over L}\int_{-L/2}^{L/2}dx \,V'(\phi) \cr &=
 {\omega(\lambda\omega^2-m^2)+ {(3\lambda\omega^2-m^2)\over L}
 \int_{-L/2}^{L/2} dx  \,\varphi +  {\lambda\over L}
\int_{-L/2}^{L/2}\,dx \,\Bigl [3\omega\varphi^2+\varphi^3 \Bigr
]=0}\,}\eqno(2)$$

\noindent Here, for concreteness  we take
$\,V(\phi)=-(1/2)m^2 {\phi^2}
+({\lambda/4})\phi^{4}+const.\,$,  $\lambda> 0$. The canonical null plane
Hamiltonian is found to be

$$ H^{l.f.}\,\equiv\int_{-L/2}^{L/2} dx \,V(\phi)
=\int_{-L/2}^{L/2} dx \,\Bigl [\omega(\lambda\omega^2-m^2)\varphi+
{1\over 2}(3\lambda\omega^2-m^2)\varphi^2+
\lambda\omega\varphi^3+{\lambda\over 4}\varphi^4
+const.\Bigr ]\,\eqno(3)$$

\noindent It is then clear form (2) and (3)
that the elimination of $\omega$ would lead
to a nonlocal Hamiltonian in contrast to the simple polynomial one obtained
if we had ignored altogether the background field variable. This seems to be
the price to pay for working on the null plane with
the corresponding simple light-front vacuum. In
the {\it continuum formulation}
the field $\,\varphi\,$ is assumed to be
an ordinary absolutely integrable function of $x$ such that
its Fourier (series) transform $\,\tilde\varphi(k,\tau)\,$
  exists together with the
inverse transform. In this case $\,{\cal C}$ vanishes
since $\,\varphi\to 0\,$ for
$\,\vert x\vert\to\infty\,$ and the constraint eq. (3), with $L\to\infty $,
follows directly on integrating
the lagrange eq. for $\varphi $.

We remind that in the context of equal-time quantization the
criterion, $\,V'(\omega)=0\, $, obtained by minimizing the classical
Hamiltonian, which determines the allowed values of the
background field,
plays a significant role in the tree level description of ssb.
In the null plane case
we do not have any physical considerations to minimize the corresponding
light-front Hamiltonian. The same criterion, however, is seen to follow
in the continuum limit now from
the constraint eq. (2) considered at the tree level. In fact
the integrals over $\,\varphi\,$,
$\,\varphi^2\,$, and $\,\varphi^3\,$ are convergent
(from the Fourier transform theory)
and the corresponding terms drop out when  $L\to\infty $.
In the renormalized theory with field operators
the value of the background field
does get modified due to the high order quantum corrections.
For the sake of the compactness we give the discussion
below directly in the continuum [9] where also the variable ${\cal C}$ is
absent.
The relevant expressions obtained
have their counterpart in the {\it discretized version} [8]
and easily identified
if we define the space integral over $[-\infty,\infty]$
by the Cauchy principal value etc.

Having discussed the Lagrangian formulation we
next construct the hamiltonian formulation following the Dirac method
which would latter be quantized. The Lagrangian density
may be rewritten as ${\cal L}={\dot\varphi}
{\varphi}^\prime-V(\phi),\,$ and it describes a
constrained or singular Lagrangian. We would treat both
 $\omega $ and $\varphi $ as canonical variables on the null-plane
and let the {\it self-consistency} requirement [5] {\it determine}
their properties.
Indicating by $p$ and $\pi$ the momenta conjugate to
$\omega $ and $\varphi $, respectively, the primary constraints
are $\,p(\tau)\approx 0\,$ and
$\,\Phi\equiv \pi-\varphi^\prime\approx 0\,$ while the canonical Hamiltonian
density is derived to be $\,{\cal H}_{c}= V(\phi)\,$ and  ${\approx}$
stands for the weak equality [5].
We postulate now the standard Poisson brackets at equal
$\tau$, with the nonvanishing brackets satisfying,
$\{p,\omega\}=-1, \;\{\pi(x),\varphi(x)\}=-\delta(x-y)$,
and assume for the preliminary Hamiltonian the expression

$$H^\prime(\tau) = {H_c}(\tau) + {\mu(\tau)}p(\tau)
                        + \int dy\; u(\tau,y)\Phi(\tau,y),\eqno(4)$$

\noindent where $\mu$ and $u$ are Lagrange multipliers.
Using (4) we derive

$${\dot p}\;=\;\{p,H^\prime\}\;{\approx}\;-\int dx
{\; V^\prime(\phi)}\;
\equiv \;-\beta(\tau),\eqno(5)$$

$${\dot\Phi}\;=\;\{\Phi,H^\prime\}\;{\approx}\;
-{\, V'(\phi)}\,-\,2u^\prime.\eqno(6)$$

\noindent The persistency requirement, $\;\dot p\approx 0\;$, then
leads to a {\it secondary constraint}
$\;\beta\approx0\;$,
while $\;\dot\Phi\approx 0\;$ results in a consistency
condition involving the multiplier $u$
and does not generate a
new constraint. An extended Hamiltonian $\,H''\,$ is now defined
by adding a term $\;\nu(\tau)\beta(\tau)\;$ to $H'$ and we repeat
the procedure. For the choice $\nu\approx 0$ no more
constraints are generated.

The three constraints $\;{p\approx0,\; \beta\approx0,
\;\Phi\approx0}$ are verified to be second class [5].
 The constraints  may be  implemented [5] by
defining modified brackets to replace the standard ones. In view of
$\;\;\{\beta(\tau),p(\tau)\}\equiv \alpha(\tau)=\int dx \;{V''(\phi)}\;$,
$\{\beta,\beta\}=\{p,p\}=0\;$ we construct the bracket $\,\{,\}^*\,$

$$\{f(x),g(y)\}^*=\{f(x),g(y)\}-{1\over{\alpha}}[\{f(x),p\}\{\beta,g(y)\}
-(\beta\leftrightarrow {p})]\,,\eqno(7)$$

\noindent with the property $\;\{f,p\}^*\;=\;\{f,\beta\}^*\;=0\;$ for
any arbitrary functional $f$. We may then set $p=0$ and $\beta=0$ even
inside these brackets and treat them as strong equalities.
It is seen
from (7) that among the surviving variables
{\it only} the bracket $\;\{\omega,\pi\}^*=\{\omega,\Phi\}^*=
\,-{\alpha^{-1}}\,{V''(\phi)}\;$ differs form
their corresponding Poisson brackets.
The remaining constraint $\Phi\approx 0$ may next be taken care of
by a modification of the brackets $\,\{,\}^*\,$ themselves
obtaining the
(final) Dirac bracket which implements all the constraints.
We check that $\,\Phi(x)\,$ is second class by itself and
$\;\{\Phi(x),\Phi(y)\}^*=\;\{\Phi(x),\Phi(y)\,\}=\,
-2{\partial_x} \delta(x-y)\;\equiv C(x,y)=
-C(y,x)$. Its (unique) inverse with the correct symmetry property
is $C^{-1}(x,y)=-C^{-1}(y,x)=\,-{\epsilon(x-y)}/4\;$ and hence
the {\it final Dirac bracket} $\,\{,\}_D\,$ is constructed as

$$\{f(x),g(y)\}_D=\{f(x),g(y)\}^* + {1\over4}\int\int dudv \{f(x),\Phi(u)\}^*
\epsilon(u-v)\{\Phi(v),g(y)\}^*.\eqno(8)$$

\noindent Inside $\,\{f,g\}_{D}\,$ we are allowed to set in addition
 $\;\pi=\varphi^\prime\;$
so that $\,\pi\,$ and $\,p\,$ are removed from the theory and we are left with
only the variables $\omega$ and $\varphi$ which are related through
the {\it constraint} $\beta=0 $ which is the same
as we found at the Lagrangian level assuring us of the
self-consistency.

{}From (8) we derive for the field $\varphi$

$$\; \{\varphi(x),\varphi(y)\}_D=-(1/4)\epsilon(x-y)\;,\eqno(9)$$

\noindent which corresponds to the well known light-cone
commutator (see below). We emphasize
that in its derivation $\{\omega,\pi\}^*$  is not required.
We also find

$$\; \{\omega,\pi(x)\}_D=\,\{\omega,\varphi'(x)\}_D=\,
{1\over2}\{\omega,\pi(x)\}^{*},\eqno(10)$$

\noindent and $\, \{\omega,\omega\}_D=0\,$ follows from symmetry
considerations. The eqs. of motion are given by $\dot f\,=\,\{f,H_{c}\}_{D}
+{{\partial f}/ {\partial\tau}} \,$
where $\,H_c(\tau)\equiv H^{l.f.}\, $ as given in (3).
For the potential considered the explicit expression for $\,\alpha\,$
reads as

$$\,\alpha(\tau)=\int dx \,V''(\phi)=
L\,(3\lambda\omega^2-m^2)\,+6\lambda \omega \,\int dx
\,\varphi\,+3\lambda
\,\int\,dx\,{\varphi^2},\eqno(11)$$

\noindent where $L\to\infty$ as discussed above.
At the {\it tree level} as discussed above the constraint implies
$\;V^\prime(\omega)=(\lambda\omega^2-m^2)\omega=0\,$ which determines the
allowed values of $\,\omega\,$. Corresponding to these values (11) shows
that $\,\alpha\to\infty\,$ and consequently $\,\{\omega,\pi\}^{*}\,=-
{\alpha}^{-1}\, V''(\phi)\to 0\,$ which from (10) leads
in the continuum limit to $\,\{\omega,\varphi(x)\}_D=0\,$.
We then find $\dot \omega=0$
which is consistent with the constant
values found for $\omega $ (from $\,V'(\omega)=0\,$) and the Lagrange
eq. of motion for $\varphi$ is also recovered. We are thus able to
build a self-consistent hamiltonian formulation in the continuum
with the separation proposed
above based on physical considerations.

The quantized theory is now obtained by
the correspondence [5] $\;i\{f,g\}_D\rightarrow
[f,g]\;$ where the quantities inside the commutator are the corresponding
quantized operators. The operator $\omega$ commutes with itself and
with the nonzero modes and no operator ordering problem
arises in contrast to the case of discretized finite volume formulation.
For the correct sign for the mass term
$\omega$ is vanishing and the self-consistency may also be checked.
In the case of the massless theory ($\lambda\ne 0$), we find from (2)
that $\omega $ is vanishing at the tree level and  from (11) we
conclude that
$\,lim_{L\to\infty}\alpha^{-1}\ne 0\,$. Consequently from (10)
$\{\omega,\varphi'\}_{D}=
\,-{\alpha^{-1}}\,{V''(\phi)}/2\;$ is nonvanishing indicating
an inconsistency and we are also unable to
recover the Lagrange eqs. of motion. This is in agreement with
the discussion in the corresponding  equal-time case where
a massless scalar theory in two dimensions is known to be ill-defined;
this is remedied (Sec. 3) by the extra space dimension
available in higher space-time dimensions.

The commutation relations of $\varphi$
may be realized in momentum space through the expansion ($\tau=0$)

$$\varphi(x)= {1\over {\sqrt{2\pi}}}\int_{-\infty}^{\infty} dk\;
{\theta(k)\over {\sqrt{2k}}}\;
[a(k)e^{-ikx}+{a^{\dag}}(k)e^{ikx}]\,\eqno(12)$$

\noindent where $a(k)$ and ${a^{\dag}}(k)$
satisfy the canonical commutation relations,
viz, $[a(k),{a(k^\prime)}^{\dag}]=\delta(k-k^\prime)$,
$[a(k),a(k^\prime)]=0$, and $[{a(k)}^{\dag},{a(k^\prime)}^{\dag}]=0$
while $\omega$ commutes with them becoming essentially a c-number.
The vacuum state is defined
by  $\,a(k){\vert vac\rangle}=0\,$,
$k> 0$.
The longitudinal momentum operator is
$\int dx:\varphi'{^2}: $ and the light-front energy
is $P^{-}=H=\int dx :V(\phi): $ where we normal order with
respect to the creation and destruction operators to drop
unphysical infinities
and we find $\,[a(k),P^{+}]=k\,a(k)\,$,
The values of $\omega=
\,{\langle\vert \phi\vert\rangle}_{vac}\;$
{\it characterize} the (non-perturbative) vacua and
the Fock space built as usual. A self-consistent
Hamiltonian formulation can thus be built in the
continuum which also can describe ssb.
The high order corrections to the (renormalized) constraint eq. (2)
will alter the tree level values of $\omega$ since, we {\it do not
have any physical considerations to normal order the
constraint equation} like we have for light-front energy or the
momentum operators. The phase transition in two dimensions can be
described [10] in the renormalized theory based on (2) and (3)
in the continuum. In the
discretized formulation we do have to face the difficult problem of
operator ordering of $\omega$ with the non zero modes
apart from making a self-consistency check as well.

\bigskip

\noindent {\bf 3. Spontaneous continuous symmetry breaking mechanism:}

Extending the discussion to continuous symmetry in $3+1$ dimensions,
the Lgrangian
density with a global {\it isospin symmetry} may be written as

$${\cal L}=[\;{\dot\varphi_a}{\varphi'}_a-
{1\over 2}(\partial_i\phi_a)(\partial_i\phi_a)-V(\phi)\;].\eqno(13)$$

\noindent Here the real scalar fields
$\;\phi_a $, $a=1,2...\;$ are the components
of an isospin-multiplet, $\,i=1,2\,$ and $\bar x\equiv(x^1,x^2)$
refer to the transverse directions,
$\;{V'_a}\equiv \delta V(\phi)/\delta{\phi}_a$, and
we have set now
$\phi_{a}(\tau,x,\bar x)=\varphi_{a}(\tau,x,\bar x)+
\omega_{a}(\tau,\bar x)$ which may be justified as before by observing
that the $\bar x$ coordinates, in contrast $x$
behave as parameters on the null-plane. We note in this connection that
for the case of free field if $\varphi(\tau,x,\bar x)$ solves the eq. of
motion so does $\varphi(\tau,x,\bar x)+\omega(\tau,\bar x)$ where
$(\,\partial_i\partial_i-m^2) \omega(\tau,\bar x)\,=0\,$.
The canonical (light-front) Hamiltonian is

$$H_{c}(\tau)=\int\,dx d^2x \Bigl [V(\phi)+{1\over 2}
(\partial_{i}\phi_{a})(\partial_{i}\phi_{a})\Bigr],\,\eqno(14)$$

\noindent and the bracket $\{,\}^*$ which implements the constraints
$\beta_{a}(\tau,{\bar x})\approx 0,\,p_a(\tau,{\bar x})\approx 0$
is now

$$\{f,g\}^*\,=\{f,g\}-\int \int d^{2}{\bar u} d^{2}{\bar v}\,
\Bigl[\,\{f,p_a({\bar u})\}C^{-1}_{ab}({\bar u},{\bar v})
\{\beta_b(\bar v),g\}-(\beta\leftrightarrow p)\;
\Bigr],\eqno(15)$$

\noindent where $C^{-1}$ is the inverse of the matrix (suppressing $\tau$)

$$C_{ab}({\bar x},{\bar y})\equiv
\{\beta_{a}({\bar x}),p_b({\bar y})\}\,
=\Bigl[\,L[-\,\delta_{ab}\,\partial_i\partial_i
+V''_{ab}(\omega)\,]+V'''_{abc}(\omega)\int dx\,
\varphi_{c}+...\Bigr]\,\delta^{2}({\bar x}-{\bar y}).\eqno(16)$$

\noindent Again
$\{\,\omega_{a}(\bar x),\pi_{b}(y,\bar y)\,\}^*= [V''_{ac}(\phi(y,\bar
y))-\delta_{ac}{\partial^{\bar y}}_{i}{\partial^{\bar y}}
_{i}]C^{-1}_{cb}(\bar x,\bar y)$ are
the only ones which differ from
the corresponding standard Poisson brackets among the surviving
variables. The {\it final} Dirac bracket which implements also the,
$\,\Phi_a(\tau,x,\bar x)\equiv \pi_{a}-{\varphi'}_{a} \approx 0 $, is

$$\{f,g\}_D\,=\,\{f,g\}^{*}\,+{1\over {4}}\int \int d^3u d^3v \,\{f,
\Phi_a(u,\bar u)\}^{*}\,{\epsilon(u-v)}\delta^{2}({\bar u}-{\bar v})\,\{\Phi_a
(v,\bar v),g\}^{*}, \eqno(17)$$

\noindent and we find (without requiring to use
$\{\omega_a,\pi_b\}^{*}$)

$$\{\,\varphi_a(x,\bar x),\varphi_b(y,\bar y)\,\}_D=
-{1\over 4}\delta_{ab}\,\epsilon(x-y)\delta^{2}({\bar x}-{\bar y})
\eqno(18),$$


A Taylor expansion in the constraint $\beta_a =0$ gives

$$L\,[\,V'_a(\omega)-\partial_i\partial_i \omega_a\,]+
\,V''_{ab}(\omega)\int dx \varphi_{b}+\,{1\over {2!}}
V'''_{abc}(\omega)\int dx \varphi_b\varphi_c+...=0, \eqno(19)$$

\noindent and at the tree level it leads to
$[\,V'_a(\omega)-\partial_i\partial_i \omega_a\,]=0\,$. As before
this equation and (19) are in agreement with the Lagrangian
formulation. The (classical level) constant
solutions for $\omega_a \,$ are determined by solving
$\,V'_{a}(\omega)=0\,$ and are relevant for
describing the ground state and ssb.
The $\bar x$-dependent solutions correspond to the solitary waves
of the equal-time formulation but here only in three or more dimensions.

For the other brackets we find

$$\,\{\omega_{a}(\bar x),\pi_{b}(y,\bar y)\}_{D}\,=\,
\{\omega_{a}(\bar x),\varphi'_{b}(y,\bar y)\}_{D}\,=\,{1\over 2}
\{\omega_{a}(\bar x),\pi_{b}(y,\bar y)\}^{*}\,\eqno(20)$$

$$\{\omega_{a}(\bar x),\omega_{b}(\bar y)\}_D\,=\,{1\over {4}}\int \int
d^{3}u d^{3}v
 \,\{\omega_{a}(\bar x),\pi_c(u,\bar u)\}^{*}\,{\epsilon(u-v)}\delta^{2}
({\bar u}-{\bar v})\,\{\pi_c
(v,\bar v),\omega_{b}(\bar y)\}^{*}. \eqno(21)$$

\noindent We discuss now the inverse matrix $C^{-1}$ which is needed to
implement the constraint (19).
For the potential
$V(\phi)={ \,{(\lambda/4)} (\phi_a\phi_a-m^2/\lambda)^2}\,$,
$\lambda> 0 $,
the constant solutions
are found from  $\,{V'}_{a}(\omega)=\,(\lambda \omega^2
-m^2)\,\omega_{a}=0\,$, where $\omega^2\equiv \omega_{a}\omega_{a}$.
In the broken symmetry phase $\,\omega^2=
(m^2/\lambda)\,$ while in
the symmetric phase (or when the potential has
the correct sign for the mass term )
$\,\omega_{a}=0$.
In the latter case
the leading term in (16) is, $\,-L(\partial_{i}\partial_{i}
+m^2)\,\delta_{ab}\delta^{2}({\bar x}-
{\bar y})\,$, while in the former it is
$\,L\,[-\delta_{ab}\partial_{i}\partial_{i}
+2m^2\,P_{ab}\,]\delta^{2}({\bar x}-{\bar y})\,$
where $\,P_{ab}=(\omega_{a}\omega_{b})/\omega^{2}\;$ is a
projection operator. The corresponding inverse matrices contain an explicit
factor $1/L$ multiplying a Green's function which is well defined.
Hence in the $L\to\infty$ limit the
$\,\{\omega_{a},\pi_{b}\}^{*}\,$ are vanishing and from (20)
and (21) we find $\,\{\omega_{a}(\bar x),\omega_{b}(\bar y)
\}_{D}=0$ and $\;\{\omega_{a}(\bar x),\varphi_{b}(y,\bar y)\}_{D}=0\,$. The
self-consistency is verified like in Sec. 2.
The transverse
directions now present cure also the inconsistency encountered in the
massless theory in two dimensions.
We can also give a new demonstration
of Coleman's theorem [11] on
the absence of Goldstone bosons in two dimensions.
There are no transverse directions on the null-plane in two dimensions
and the matrix  $\,C_{ab}=2Lm^2 P_{ab}\,$ contains a projection operator
which can not be inverted
even when $m\ne 0$. We are unable to implement the constraints  and
construct a self-consistent hamiltonian formulation.
The Fock space operators now depend on the
transverse momentum as well and they satisfy
$[a_{b}(k,\bar k),{a_{c}(k',{\bar k'})}^{\dag}]=\delta_{bc}
\delta(k-k')\delta^{2}({\bar k}-{\bar k'})\,$  etc. where
$\,{\bar k}=(k^{1},k^{2})\,$ indicates the transverse components. The
quantized theory is also checked to be self-consistent (and even for
the case of $\bar x$-dependent solutions which we do not consider for our
purpose).

The global invariance of (13) at the {\it classical} level gives rise to
conserved isospin currents and the field theory
generators are given by

$$G_{\alpha}(\tau)=\, \int d^{2}{\bar x}\,p({\bar x})
\,t_{\alpha}\,
\omega({\bar x})\, +\int dxd^{2}{\bar x}\,\pi(x,{\bar x})\,t_{\alpha}
\,\varphi (x,{\bar x})\,\eqno(22)$$

\noindent where $\alpha,\beta $ are the group indices,
$t_{\alpha}$ are hermitian and antisymmetric,
and  $\,[t_{\alpha},t_{\beta}\,]=\,
if_{\alpha\beta\gamma}\, t_{\gamma}$. The
 generators in the {\it quantized theory} are hence
given by $(p_a=0,\,\pi_{a}={\varphi'}_{a}\,)$

$$G_{\alpha}(\tau)=\,- i \int d^{2}{\bar x}\,
 dx\,{{\varphi}'}_{a}(x,{\bar x})\,(t_{\alpha})_{ab}\,\varphi_{b}(x,{\bar x})
\,=\,\int d^{2}{\bar k}\,dk \, \theta(k) {a_{a}(k,{\bar k})
^{\dag}}\,(t_{\alpha})_{ab}\,a_{b}(k,{\bar k})\,\eqno(23).$$

\noindent which come out normal ordered
and consequently on the null-plane the continuous symmetry generators
always  annihilate the vacuum.
We find that
$\,[G_{\alpha},\varphi_{a}\,]=\,-(t_{\alpha})_{ab}\varphi_{b}$,
$\,[G_{\alpha},\omega_{a}\,]=0\,$, and $\,[G_{\alpha},G_{\beta}\,]=\,
if_{\alpha\beta\gamma}\, G_{\gamma}$.
The tree level ssb is now
described as follows. A particular solution,
$(\omega_1,\omega_2,\omega_3...)$, of $ V'_{a}=\,\omega_{a}
(\lambda\omega^2-m^2)=0\,$
defines a preferred direction in the isospace
which {\it characterizes} a (non-perturbative)
vacuum state, $\;{\langle\vert \phi_a\vert\rangle}_{vac}=
\;\omega_a\;$.
The Fock space of the corresponding
physical sector in the quantized theory is built by applying
the particle creation operators on this vacuum state.
The degeneracy of the vacuum is described by the
continuum of the allowed orientations in the isospin space of the
background isovector.
In the symmetric phase $\,\omega_a=0\,$ and there is no
preferred direction.
In the broken phase the potential
expressed in terms of the field operators $\varphi_{a}$ and $\omega_{a}$
reveals that the surviving
symmetry in the quantized theory is of lesser
dimension than the initial one
and we obtain Goldstone bosons in the theory.
Not all the generators are now conserved in the quantized theory
but there may survive a set of linearly independent field generators which
still do so. They are
evidently found by solving
$({\tilde t}_{\alpha})_{ab} \omega_{b}=0\,$
where $\,{\tilde t}_{\alpha}\,$ are appropriate linearly independent
combinations, depending on the $\,\omega_{a}\,$,
of the original matrix generators. The corresponding operators
$\,{\tilde G}_{\alpha}\,$ generate the {\it surviving symmetry}
in the quantized theory.
The number of Goldstone bosons may be counted following the
arguments as in the case of equal-time quantization [12].
The implications of the lack of conservation at the quantum level
of some currents conserved in the classical theory needs further study.
The description of the tree level {\it Higgs mechanism}
is straightforward and like in equal-time quantization.
To give its quantized description the null plane
theory of interacting gauge field (e.g., QCD)
has to be understood first since the
constraints would contain more terms when
fermionic and other bosonic interactions are also present.

\vskip 1.5cm

\noindent{\bf Acknowledgements:}
\vskip 0.2cm
The author acknowledges gratefully
the hospitality of the Department of Physics,
Ohio State University, of the INFN and the Department of Physics,
University of Padova, and the financial support from INFN-Padova.
Acknowledgements with thanks are due to Ken
Wilson, Steve Pinsky, Stuart Raby, Antonio Bassetto,
Mario Tonin, and G. Costa for useful remarks and questions and
specially, to Robert Perry
and Avaroth Harindranath for extensive clarifications which were
indispensable for completing this work.

\vskip 1.5cm

\noindent{\bf References:}

\item{[1.]}P.A.M. Dirac, Rev. Mod. Phys. {\bf 21} (1949) 392.
\item{[2.]}S. Weinberg, Phys. Rev. {\bf 150} (1966) 1313; J.B. Kogut and
D.E. Soper, Phys. Rev. {\bf D1} (1970) 2901.
\item{[3.]}H.C. Pauli and S.J. Brodsky, Phys. Rev. {\bf D 32} (1985) 1993
and 2001; Phys. Rev. {\bf D 32} (1985) 2001; {\it recent review}:
S.J. Brodsky and H.C. Pauli, {\it Schladming Lectures}.
SLAC preprint SLAC-PUB-5558/91 and the references therein.
\item{[4.]}K.G. Wilson, Nucl. Phys. B (proc. Suppl.) {\bf 17} (1990);
R.J. Perry, A. Harindranath, and K.G. Wilson,
Phys. Rev. Lett. {\bf 65} (1990)  2959.
\item{[5.]}P.A.M. Dirac, {\it Lectures
in Quantum Mechanics}, Benjamin, New York, 1964. See also
E.C.G. Sudarshan and
N. Mukunda, {\it Classical Dynamics: a modern perspective}, Wiley, N.Y.,
1974; A. Hanson, T. Regge and C. Teitelboim, {\it Constrained
Hamiltonian Systems}, Acc. Naz. dei Lincei, Roma, 1976.
\item{[6.]}T. Maskawa and K. Yamawaki, Prog. Theor. Phys. {\bf 56} (1976)
270; N. Nakanishi and K. Yamawaki, Nucl. Phys. {\bf B122} (1977) 15;
R.S. Wittman, in Nuclear and Particle Physics on the Light-cone, eds.
M.B. Johnson and L.S. Kisslinger, World Scientific, Singapore, 1989.
\item{[7.]}Th. Heinzl, St. Krusche, and E. Werner, Regensburg preprint TPR
91-23, Phys. Lett. {\bf B 256} (1991) 55; {\bf B 272} (1991) 54.
\item{[8.]}P.P. Srivastava, {\it Spontaneous symmetry breaking mechanism
in light-front quantized field theory- Discretized formulation},
preprint, Ohio State University 92-0173 (Slac PPF-9222), April 1992.
\item{[9.]}The separation of the background field and  its treatment
as a dynamical variable were
suggested in Ohio State preprints 91-0481 (Slac PPF-9148) and 92-0012
(Slac PPF-9202),  November and December
91, on {\it  Spontaneous symmetry breaking} and {\it
Higgs mechanism} in {\it light-front quantized field
theory} respectively.
\item{[10.]}{\it Light-front field theory and nature of phase transition in
$({\phi^{4}})_{2}$ theory}, Padova university preprint DFPF/93/TH/18, March
1993.
\item{[11.]}S. Coleman, Commun. Math. Phys. {\bf 31} (1973) 259.
\item{[12.]}S. Weinberg, Phys. Rev. Lett.  {\bf 29} (1972) 1698.

\bye